%
\let\includefigures=\iftrue
%
%
%
%
%
\input harvmac
\input rotate
\input epsf
\input xyv2
\noblackbox
\includefigures
\message{If you do not have epsf.tex (to include figures),}
\message{change the option at the top of the tex file.}
\def\figin{\epsfcheck\figin}\def\figins{\epsfcheck\figins}
\def\epsfcheck{\ifx\epsfbox\UnDeFiNeD
\message{(NO epsf.tex, FIGURES WILL BE IGNORED)}
\gdef\figin##1{\vskip2in}\gdef\figins##1{\hskip.5in}
\else\message{(FIGURES WILL BE INCLUDED)}%
\gdef\figin##1{##1}\gdef\figins##1{##1}\fi}
\def\DefWarn#1{}

\def\figinsert{\goodbreak\midinsert}
\def\ifig#1#2#3{\DefWarn#1\xdef#1{fig.~\the\figno}
\writedef{#1\leftbracket fig.\noexpand~\the\figno}%
\figinsert\figin{\centerline{#3}}\medskip\centerline{\vbox{\baselineskip12pt
\advance\hsize by -1truein\noindent\footnotefont{\bf
Fig.~\the\figno:} #2}}
\bigskip\endinsert\global\advance\figno by1}
\else
\def\ifig#1#2#3{\xdef#1{fig.~\the\figno}
\writedef{#1\leftbracket fig.\noexpand~\the\figno}%
\global\advance\figno by1} \fi
\def\Title#1#2{\rightline{#1}\ifx\answ\bigans\nopagenumbers\pageno0
\else\pageno1\vskip.5in\fi \centerline{\titlefont #2}\vskip .3in}
\font\caps=cmcsc10

\def\yboxit#1#2{\vbox{\hrule height #1 \hbox{\vrule width #1
\vbox{#2}\vrule width #1 }\hrule height #1 }}
\def\fillbox#1{\hbox to #1{\vbox to #1{\vfil}\hfil}}
\def\ybox{{\lower 1.3pt \yboxit{0.4pt}{\fillbox{8pt}}\hskip-0.2pt}}

\def\rightarrowbox#1#2{
  \setbox1=\hbox{\kern#1{${ #2}$}\kern#1}
  \,\vbox{\offinterlineskip\hbox to\wd1{\hfil\copy1\hfil}
    \kern 3pt\hbox to\wd1{\rightarrowfill}}}

\def\CO{{\cal O}}

\def\tilde{\widetilde}

           \def\CO{{\cal O}}


\def\dj{\hbox{d\kern-0.347em \vrule width 0.3em height 1.252ex depth
-1.21ex \kern 0.051em}}

\def\pt{\partial}

\def\Dirac{\,\raise.15ex\hbox{/}\mkern-13.5mu D}
\def\dirac{\,\raise.15ex\hbox{/}\kern-.57em \partial}
\def\aslash{\,\raise.15ex\hbox{/}\mkern-13.5mu A}

\def\shalf{{\ifinner {\textstyle {1 \over 2}}\else {1 \over 2} \fi}}
\def\sshalf{{\ifinner {\scriptstyle {1 \over 2}}\else {1 \over 2} \fi}}
\def\sfourth{{\ifinner {\textstyle {1 \over 4}}\else {1 \over 4} \fi}}
\def\sthreehalfs{{\ifinner {\textstyle {3 \over 2}}\else {3 \over 2} \fi}}
\def\sdhalfs{{\ifinner {\textstyle {d \over 2}}\else {d \over 2} \fi}}
\def\sdmtwohalfs{{\ifinner {\textstyle {d-2 \over 2}}\else {d-2 \over 2} \fi}}
\def\sdmasonehalfs{{\ifinner {\textstyle {d+1 \over 2}}\else {d+1 \over 2} \fi}}
\def\sdmasthreehalfs{{\ifinner {\textstyle {d+3 \over 2}}\else {d+3 \over 2} \fi}}
\def\sdmastwohalfs{{\ifinner {\textstyle {d+2 \over 2}}\else {d+2 \over 2} \fi}}


\lref\Hayden{
  P.~Hayden and J.~Preskill,
  ``Black holes as mirrors: quantum information in random subsystems,''
  JHEP {\bf 0709}, 120 (2007)
  [arXiv:0708.4025 [hep-th]].
}

\lref\Sekino{
  Y.~Sekino and L.~Susskind,
  ``Fast Scramblers,''
  JHEP {\bf 0810}, 065 (2008)
  [arXiv:0808.2096 [hep-th]].
    L.~Susskind,
  ``Addendum to Fast Scramblers,''
  arXiv:1101.6048 [hep-th].
}

\lref\Susskindnew{
  L.~Susskind,
  ``Addendum to Fast Scramblers,''
  arXiv:1101.6048 [hep-th].
}

\lref\Damour{
  T.~Damour,
  ``Black Hole Eddy Currents,''
  Phys.\ Rev.\  D {\bf 18}, 3598 (1978).
}

\lref\Thorne{
  K.~S.~.~Thorne, R.~H.~.~Price and D.~A.~.~Macdonald,
  ``Black Holes: The Membrane Paradigm,''
{\it  New Haven, USA: Yale Univ. Pr. (1986) 367p}
}

\lref\Susskindbook{
  L.~Susskind and J.~Lindesay,
  ``An introduction to black holes, information and the string theory
  revolution: The holographic universe,''
{\it  Hackensack, USA: World Scientific (2005) 183 p}
}

\lref\brick{
  G.~'t Hooft,
  ``On the Quantum Structure of a Black Hole,''
Nucl.\ Phys.\ B {\bf 256}, 727 (1985).
}

\lref\tesselations{
R.~Mosseri and J.~F.~Sadoc,
``The Bethe lattice: a regular tiling of the hyperbolic plane",
J. Physique - Lettres {\bf 43}, 249 (1982) 

B. S\"oderberg,
``Bethe Lattices in Hyperbolic Space",
Phys. Rev. E {\bf 47}, 4582 (1993).
}

\lref\expander{
  J.~L.~F.~Barbon and J.~M.~Magan,
  ``Fast Scramblers, Horizons and Expander Graphs,''
JHEP {\bf 1208}, 016 (2012).
[arXiv:1204.6435 [hep-th]].
}
\lref\small{
  J.~L.~F.~Barbon and J.~M.~Magan,
  ``Fast Scramblers Of Small Size,''
JHEP {\bf 1110}, 035 (2011).
[arXiv:1106.4786 [hep-th]].
}
\lref\caos{
  J.~L.~F.~Barbon and J.~M.~Magan,
  ``Chaotic Fast Scrambling At Black Holes,''
Phys.\ Rev.\ D {\bf 84}, 106012 (2011).
[arXiv:1105.2581 [hep-th]].
}

 \lref\expanderg{
S. Hoory, N. Linial and A. Widgerson,
``Expander graphs and their applications."
Bulletin of the American Mathematical Society, {\bf 43} 4, 439 (2006)

A. Lubotzky,
``Expander Graphs in Pure and Applied Mathematics",
arXiv:1105.2389 [math.CO].
}

\lref\lashkari{ N.~Lashkari, D.~Stanford, M.~Hastings, T.~Osborne and P.~Hayden,
  ``Towards the Fast Scrambling Conjecture,''
JHEP {\bf 1304}, 022 (2013).
[arXiv:1111.6580 [hep-th]].   }

\lref\berenstein{ C.~Asplund, D.~Berenstein and D.~Trancanelli,
  ``Evidence for fast thermalization in the plane-wave matrix model,''
  Phys. Rev. Lett. 107, 171602 (2011)
  arXiv:1104.5469 [hep-th].
  
   C.~Asplund, D.~Berenstein and E.~ Dzienkowki,
   ``Large N classical dynamics of holographic matrix models,''
     Phys. Rev. D, 87, 084044 (2013)
     arXiv:1211.3425 [hep-th].
}

\lref\fischler{M.~Edalati, W.~Fischler, J.~F.~Pedraza and W.~T.~Garcia,
  ``Fast Scramblers and Non-commutative Gauge Theories,''
  JHEP {\bf 1207}, 043 (2012).
  arXiv:1204.5748 [hep-th].}

\lref\brady{L.~Brady and V. Sahakian,
  ``Scrambling with matrix black holes,''
  Phys. Rev. D 88, 046003 (2013).
  arXiv:1306.5200 [hep-th].}

\lref\shenker{S.~Shenker and D.~Stanford,
  ``Black holes and the butterfly effect,''
  arXiv:1306.0622 [hep-th].}

\lref\axenides{M.~Axenides, E.~G.~Floratos and S.~Nicolis,
      ``Modular discretization of the AdS2/CFT1 Holography,''
      arXiv:1306.5670 [hep-th].}

\lref\cesar{G.~Dvali, D.~Flassig, C.~Gomez, A.~Pritzel and N.~Wintergest,
  ``Scrambling in the Black Hole Portrait,''
  arXiv:1307.3458 [hep-th].}

 \lref\zabrodin{
A. V. Zabrodin,
``Non-Archimidean strings and Bruhat-Tits trees."
Commun. Math. phys. 123, 463-483 (1989).

}

\lref\virasoro{
  R.~Rammal, G.~Toulouse and M.~A.~Virasoro,
  ``Ultrametricity for physicists,''
Rev.\ Mod.\ Phys.\  {\bf 58}, 765 (1986).
}

\lref\ultradiff{A.~T.~ Ogielski and D.~L.~Stein, ``Dynamics on Ultrametric Spaces", Phys. Rev. Lett. {\bf 55}, 1634 (1985).

G. Paladin, M. M\'ezard and C. de Dominicis, ``Diffusion in an ultrametric space: a simple case". J. Physique Lett {\bf 46}, 985 (1985).

  C.~Bachas and B.~A.~Huberman,
  ``Complexity And Ultradiffusion,''
Phys.\ Rev.\ Lett.\  {\bf 57}, 1965 (1986).
ibid J.\ Phys.\ A {\bf 20}, 4995 (1987).
}



\line{\hfill IFT UAM/CSIC-13-066}

\vskip 0.7cm

\Title{\vbox{\baselineskip 12pt\hbox{}
 }}
{\vbox {\centerline{ Fast Scramblers 
 }
\vskip10pt
\centerline{ And}
\vskip10pt
\centerline{Ultrametric Black Hole Horizons}
}}
\vskip 0.5cm

\centerline{$\quad$ {
{\caps Jos\'e L.F. Barb\'on}$^{\dagger}$
 {\caps and}
{\caps Javier M. Mag\'an}$^{\star}$
}}
\vskip0.5cm

\centerline{{\sl  $^\dagger$ Instituto de F\'{\i}sica Te\'orica IFT UAM/CSIC }}
\centerline{{\sl 
 UAM, Cantoblanco 28049. Madrid, Spain }}
\centerline{{\tt jose.barbon@uam.es}}

\vskip0.2cm

\centerline{{\sl $^\star$
International Institute of Physics}}
\centerline{{\sl  Universidade Federal do Rio Grande do Norte}}
\centerline{{\sl  59012-970 Natal, Brazil}}
\centerline{{\tt javier.martinez@uam.es}}

\vskip1.2cm

\centerline{\bf ABSTRACT}

 \vskip 0.3cm

 \noindent

We propose that fast scrambling on finite-entropy stretched horizons can be modeled by a diffusion process on an
effective ultrametric geometry. A scrambling time scaling logarithmically with the entropy is obtained when the elementary transition rates saturate causality bounds on the stretched horizon. The so-defined ultrametric diffusion becomes unstable in the infinite-entropy limit. A  formally regularized version can be shown to follow a particular case of the  Kohlrausch law.

\vskip 0.2cm

\Date{October  2013}

\vfill





\baselineskip=15pt

\newsec{Introduction}

\noindent

The notion of stretched horizon (SH) has been a recurring meme  in the study of black hole dynamics \refs{\Damour, \Thorne, \Susskindbook}. 
Physical processes in  the near-horizon region effectively freeze out if  measured in terms of the asymptotic static time variable. In describing the interactions of a black hole with its surroundings,  it becomes then useful to replace the intricate near-horizon details by an effective system defined on a time-like surrogate of the horizon. At the classical level this picture is known as the `membrane paradigm', which is very efficient in modeling the astrophysics of realistic black holes, and specifies the SH as a hydrodynamical system with a set of characteristic transport coefficients. At the quantum level, a finite region  of the SH with  area $A$  is specified to have a finite density of states proportional to $\exp(S_A) = \exp(A/4G) $ where $G $ is Newton's constant. Matching to the near-horizon description by local quantum field theory (technically defined as an expansion in powers of $1/S_A$)
fixes the thickness of the SH to be Planckian (cf. \refs\brick).

A systematic derivation of the SH effective theory from  {\it ab initio} holographic formulations, such as the AdS/CFT correspondence, remains elusive except in the  limit of infinite specific entropy $S_A =\infty$, in which the  remaining relevant time scale is given by the inverse Hawking temperature $\beta=T^{-1}$. Other time scales, such as the evaporation scale $t_{\rm ev} \sim \beta \,S$ and the recurrence scale $t_{\rm rec} \sim \beta \,e^S$ exist at finite $S$.  A new one was proposed in 
 \refs{\Hayden, \Sekino} associated to the  {\it fast scrambling} conjecture, stating that near-horizon thermalization processes should saturate causality constraints. On general grounds, the arguments put forward in \refs{\Hayden, \Sekino} identify the relevant time scale as the light-crossing time across the near-horizon (Rindler) region, down to the Planckian stretched horizon. With this formal characterisation, it was shown in \refs\small\ that one always obtains a universal result for the scrambling time, valid for any type of horizon:
\eqn\st{
t_S \sim \beta\,\log \,S_*\;,
}
where $\beta = T^{-1}$ is the inverse Hawking temperature and $S_*$ is the effective number of degrees of freedom in one thermal cell, i.e. the amount of horizon entropy contained in one SH volume of size $\beta$ (measured at the outer edge of the Rindler region). 

The heuristic arguments behind the original proposal  have a distinctive kinematical flavor, to the extent that the interpretation of \st\ as a scrambling time is far from obvious. In this paper we review previous work showing that \st\ does arise as the true scrambling time scale for various diffusion processes which can be shown to arise in the near-horizon region under some extreme dynamical assumptions. This is done in section 2. In section 3 we introduce a more phenomenological approach where we discuss {\it non-local} diffusion processes taking place at the SH, assuming saturation of causality constraints, as dictated by the near-horizon geometry. A natural mathematical structure encoding these constraints is the notion of {\it ultrametricity}. We end in section 3 with some conclusions and outlook.

\newsec{Optical Frame And Atmospheric Scrambling}
\noindent

Working locally, let us model a stretched horizon as the membrane (more generally $d$-brane) sitting at fixed $y=y_*$ coordinate in $(d+2)$-dimensional Rindler,
\eqn\rnd{
ds^2 = -y^2 \,dt^2 + dy^2 + d{\vec x}^{\,2}
\;,}
which approximates near $y=0$ any non-degenerate event horizon over distances small enough to neglect the local curvature. We use dimensionless coordinates in units where $\beta/2\pi=1$, with $T=1/\beta$  the Hawking temperature. The local temperature
at coordinate $y$ is then $1/2\pi y$, so that the metric must be cut-off at the end of the near-horizon region,  $y\approx 1$, and subsequently matched to the far-metric of the black object. In realistic situations, the metric \rnd\ is taken as an approximation of the near-horizon region on sufficiently small  longitudinal scales of order $|\Delta {\vec x}\,| \ll \beta$, well within   a thermal cell of the horizon. 
In order to fix $y_*$ in our units, we notice that the Bekenstein--Hawking entropy of an ${\vec x}$ domain of size $\beta$ is $S_* \sim (\beta/\ell_{\rm P})^d = (2\pi /\ell_{\rm P})^d$. Since the SH is determined by a Planckian local temperature
$$
T_* = {1\over 2\pi y_*} = m_{\rm P} = {1\over \ell_{\rm P}} \;,
$$
we find the matching
\eqn\mat{
(y_*)^d = {1\over S_*}
}
with determines $y_*$ in our dimensionless units. 

Despite the fact that the induced metric at the SH surface $y=y_*$  is flat, the causality constraint on the SH is nontrivial. One way to exhibit this fact uses a conformally related metric, the so-called `optical frame', obtained by rescaling the physical metric in such a way that the static redshift disappears, i.e. for  \rnd\ we have 
\eqn\hy{
{\widetilde ds}^{\,2} = -dt^2 + {dy^2 + d{\vec x}^{\,2} \over y^2} 
\;,}
which describes the direct product of the time line with a $(d+1)$-dimensional hyperbolic space. Being conformally related, the metrics \rnd\ and \hy\ share the same local causal structure. In particular, null paths connecting two points at the $y=y_*$ surface correspond to geodesic hyperbolic arcs in the  ${\bf H}^{d+1}$ space, cut-off at $y=y_*$. The resulting time of flight across a ${\vec x}$-coordinate domain of magnitude $2L$
 is
\eqn\tlog{
t_{L} = 2\log (L/y_*) = 2z_L
\;,}
where we have defined the so-called `tortoise' or Regge--Wheeler coordinate,
$
z=\log (y/y_*)
$,
normalized so that $z_* =0$ at the stretched horizon. Equation \tlog\ shows that the fastest causal connection between two points separated by an amount of order $L$ in the SH scales only logarithmically with $L$. This contrasts with the causal time for any transport {\it within} the SH which, if taken in isolation, would be proportional to $L$, or even $\CO(L^2)$ if the usual diffusion processes are assumed. 
Fast scramblers can be defined as those for which the spread of information saturates (parametrically) the causality bound. For the  SH to thermalize that fast, the relation \tlog\ implies  that a patch of $S_*$ degrees of freedom, with size $L_* \sim \ell_{\rm P} \,S_*^{1/d}$ must be scrambled in a time
$
t_S \sim 
\log (S_*)
$, 
in units of the inverse Hawking temperature $\beta$. 

As pointed out in \refs\expander, the entire dynamics of low-energy effective field theory (LEFT) on \rnd\ can be rewritten on the optical frame \hy\ by means of a conformal map which is formally singular at the horizon. The interesting fact about such an optical-frame description  is that 
the contribution of an effective operator of dimension $\Delta$ to the action in the physical frame \rnd:
\eqn\margr{
\int_{\rm Rindler}  \Lambda^{d+2 -\Delta} \,\CO_\Delta
\;,}
controlled by 
 a mass scale $\Lambda$, translates into a position-dependent potential in the optical frame \hy:
 \eqn\margo{
 \int_{{\bf R} \times {\bf H}^{d+1}} (\Lambda_{\rm eff})^{d+2-\Delta} \,{\tilde \CO}_\Delta\;,
 }
  controlled by the inhomogeneous scale  
\eqn\rsc{
\Lambda_{\rm eff} (y) \approx \Lambda \,y\;,
} 
so that relevant operators on the LEFT  are turned off as we approach the horizon, while irrelevant operators, whose physical effects scale with inverse powers of $\Lambda_{\rm eff}$, are enhanced. In particular, any
operator suppressed by $\Lambda= m_{\rm P}$ in the physical frame becomes strongly coupled on the SH, when evaluating its contribution to physical processes at energies of the order of the Hawking temperature. Thus, the optical frame gives an operational definition of the stretched horizon as the Wilsonian threshold of the LEFT, specified at the cutoff $y=y_*$ surface of ${\bf R} \times {\bf H}^{d+1}$. 

The equivalent presentation of  near-horizon dynamics in terms of the  \hy\  geometry allows us to construct some simple models of what could be termed `fast atmospheric scrambling'. Specifying a thermal state on the effective theory on  \hy\ and assuming interactions of $\CO(1)$ strength, we can model the propagation of a probe on \hy\ as a random walk with step of order $\beta$, playing out on an discretized hyperbolic space, i.e. an {\it expander graph} (cf. \refs\expander, \refs\expanderg). Standard results in expander graph theory imply that a graph of size $S_*$ is scrambled after about $\log S_*$ steps of the effective walk. 

The expander graph model has two rather clear limitations. First, it requires a large interaction rate within the whole Rindler region, i.e. a large marginal operator with $\Delta =d+2$ in \margo. Such an assumption  is quite {\it ad hoc} in the context of  black hole physics. In fact, we should be able to neglect all interactions except  gravitational ones, which  are given by irrelevant operators of the form \margo\ with $\Delta >d+2$, hence suppressed by inverse powers of $S_*$. The second problem is that  states which start highly localized at the SH boundary of the expander graph undergo slow scrambling if no extra dynamics is postulated (cf. \refs\expander). To understand this, one notices that diffusion on \hy\ has an effective drift term biasing the diffusion towards low $y$ values. Starting from the diffusion equation on ${\bf R} \times {\bf H}^{d+1}$:
$$
\pt_t \,\rho = D\,\nabla^2_{{\bf H}^{d+1}} \,\rho
$$
with diffusion coefficient $D$, we 
we can write a `radial' equation for the reduced density $f(t, y, {\vec x}\,) = y^{-d}\,  \rho(t, y, {\vec x}\,)$ with the form 
$$
\pt_t f = Dd \,\pt_z f + D \pt_z^2 f + D y_*^2 e^{2z} \,{\vec \pt}^{\,2} f \;,
$$ 
where we have transformed to the Regge--Wheeler coordinate $z=\log (y/y_*)$. 
This equation can be interpreted as a diffusion equation on the {\it flat} $(z, {\vec x}\,) $ hyperplane with an asymmetric diffusion coefficient and a drift term directed towards the low-$z$ region. Hence, 
this directed drift prevents the probability distribution to explore high $y$ values if it happens to start at $y=y_*$. Rather, the distribution stays confined on the SH, whose induced metric is essentially flat, leading to a slow-scrambling behavior.

A somewhat different set up  which evades these problems is 
the `chaotic billiard' model of the SH (cf. \refs\caos). In this construction we just regard the SH as a chaotic version of the old brick wall of 't Hooft  (cf. \refs\brick), capable of scattering a stable probe. Since the scattering cross section at the SH should be of $\CO(\beta^2)$ in the length units of the optical metric, the chaotic billiard model of a single thermal cell can be specified by a Hadamard (hyperbolic) billiard with curvature radius $\beta$, volume $S_*$ in units of the curvature radius, and thus $\CO(S_*)$ `scatterers' on its boundary. We can model the boundary scatterers as $S_*$ spheres of radius $\beta$ carved out from the boundary of the hyperboloid. For such a billiard, the randomization time of the probe is of the order of the crossing time, which is again given by $\log S_*$ in units of $\beta$. Thus, we see that the billiard model achieves fast scrambling in the right time scale, the main problem of the model being the rather unphysical character of the probe, whose interactions with the SH are restricted to a simple elastic scattering in the fully classical approximation \foot{Other approaches to the Fast Scrambling conjecture may be found in \refs{\lashkari, \berenstein, \fischler, \brady, \shenker, \axenides, \cesar}.}

\newsec{An Ultrametric Diffusion Model}

\noindent

The two extreme examples discussed in the previous paragraphs show that hyperbolic geometry holds a key to the `fast scrambling' character, while at the same time illustrating the fact that new dynamics must be postulated at the SH. 
We now take a more phenomenological attitude and try to combine the virtues of both  previous models.  We shall continue to model scrambling in terms of discrete classical diffusion, but remove the assumption of locality. Crucially, we also require the non-local diffusion kernel
to just saturate the causality requirements embodied in the relation \tlog. In other words, we shall restrict the amount of non-locality of the SH dynamics by declaring it compatible with standard locality in the near-horizon region. This assumption is natural if we are going to preserve the applicability of the LEFT down to the edge of the SH. 

 One way of representing these conditions is to regard the SH as the boundary of a cut-off expander graph, and then `integrate out' the bulk of the expander graph, retaining only the transition amplitudes between points on the boundary \foot{A rigorous treatment of this computation, for the case of an infinite Cayley tree, may be found in \refs\zabrodin.}. Since expander graphs are well-approximated by regular trees, we can simplify matters by considering a Cayley tree of fixed branching rate $b=d+1$, which provides a discretization of a cut-off hyperbolic geometry in $d+1$ dimensions \refs\tesselations.  

A single thermal cell of the SH will be modeled by $S_*$ points over which we define  a positive probability density, $\rho$,  undergoing discrete
diffusion as specified by the equation 
\eqn\diffe{
\pt_t \,\rho = W \,\rho\;,
}
where $\rho$ is a $S_*$-vector and $W$ is a $S_*\times S_*$ matrix of transition rates. Off-diagonal elements must then be positive,
$
W_{i\neq j} \geq 0
$, 
and normalization $\sum_i \rho_i = 1$ determines the (negative) diagonal elements to be 
$
W_{ii} = - \sum_{j\neq i} W_{ij}
$.

The crucial hypothesis on the matrix $W_{ij}$ is the requirement that the individual rates $W_{ij}$ between two points $i$ and $j$ of the SH be determined by the inverse of their causal time separation $t_{ij}$, where the causal times are computed as the times-of-flight of photons across the thermal atmosphere, via equation \tlog. 
Since we model the hyperbolic geometry by a tree, the $z$ coordinate appearing in \tlog\  is to be replaced by  the branching level $n$ of the tree, measured from the SH, in units of the inverse Hawking temperature.

\bigskip
\centerline{\epsfxsize=0.6\hsize\epsfbox{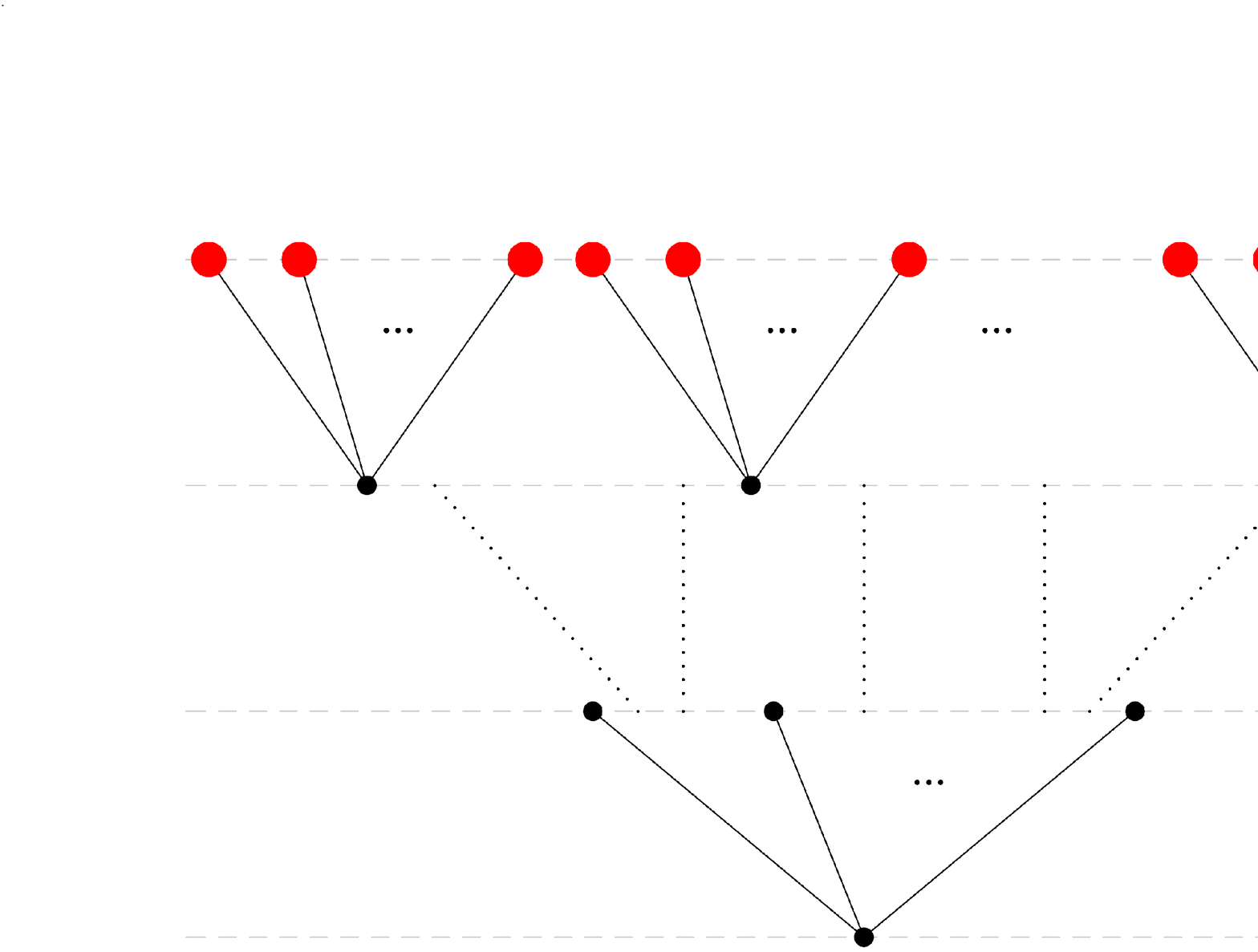}}
\noindent{\ninepoint\sl \baselineskip=2pt {\bf Figure 1:} {\ninerm
Ultrametric tree structure associated to the transition matrix on the stretched horizon (red dots). The random walk is defined just on the endpoints of the tree. 
}}
\bigskip

In order to avoid confusions with the use of expander graph diffusion in the previous section, we stress that the internal nodes of the tree are {\it not} part of the SH, nor is the diffusion process \diffe\ `playing out' on the tree. Rather, the tree is just used as a geometrical encoding of the  hierarchy of transition rates. 
We regard the points on the SH as the top leaves of an auxiliary  $b$-branching tree with $b=d+1$. Any point on the SH has a single ancestor at level $n$, which is shared among a total of $b^n = (d+1)^n$ points in the SH. The elementary transition time between any such points sharing a common ancestor is taken to be proportional to the level $n$ (by equation \tlog), and be equal for all points sharing the same ancestor. 
The characteristic time scale for the variation of the local density also depends on the phase space of final states. 
In particular, the total probability to jump from site $i$ to any other point connected through a level-$n$ ancestor is 
\eqn\trate{
\sum_{j_n} W_{ij_n} = d(d+1)^{n-1}\,W_n \equiv e^{-I_n}\;,
}
where we have defined an `effective action' for the transition through level $n$, characterizing the effective `barrier height'.  It becomes natural to include the entropic factor $d(d+1)^{n-1}$ in the relation between the characteristic time scales and the matrix of transition rates:
\eqn\relt{
e^{-I_n} \sim {1\over t_n} \sim {1\over n}\;.}
This expression is the main assumption of our model, stating that the effective action for elementary transitions has a particular dependence with the ultrametric distance between initial and final states: 
\eqn\logheights{
I_n = \log \,n + {\rm constant}\;,
}
 with exactly unit coefficient in front of the logarithm.

Such a  hierarchical structure  is characteristic of so-called {\it ultrametric}
 spaces \refs\virasoro, in which distances are defined by ancestry rules such as those implied by the regular tree specified here.
 More generally, an ultrametric structure defines distances in terms of an ancestry level on a tree (not necessarily a regular one). 
The general solution to the diffusion problem \diffe\ on ultrametric spaces is well known (see for example \refs{\ultradiff}) and has been intensely studied in the context of disordered media and spin glasses. The relevant information is encoded in the spectral properties of the transition kernel $W_{ij}$.

The completely scrambled state $\rho_i = 1/S_*$ is an eigenvector of $W$ with vanishing eigenvalue, all the rest being negative-definite. We choose to parametrize them in terms of the quasi-normal frequencies $\Gamma_r$ which are the eigenvalues of the negative kernel $-W_{ij}$ and read,
\eqn\eigenv{
\Gamma_r = \sum_{k=n_* -r+2}^{n_*} e^{-I_k} + {d+1 \over d}\,e^{-I_{n_* -r+1}} \;.
}
In this expression, the index $r$ runs over all integers in the interval $0\leq r\leq n_*$, where $n_*$ stands for the effective maximal ultrametric distance between the $S_*$ points of the SH under consideration, and is determined by the relation $(d+1)^{n_*} = S_*$. The expression \eigenv\ degenerates for $r=0, 1$, for which we have
$
\Gamma_0 = 0$ and 
$$ \Gamma_1 = \left({d+1 \over d}\right)\,e^{-I_{n_*}}\;.$$ 
Notice that $\Gamma_1$ is the leading quasi-normal frequency which, for the particular case of interest \logheights\ is given by
$
\Gamma_1 \sim 1/ \log \,S_*$, 
up to coefficients of $\CO(1)$. Being the smallest positive quasi-normal frequency, it determines the time scale for exponential relaxation to the uniform distribution, 
\eqn\ufs{
t_S \sim {1\over \Gamma_1} \sim \log\,S_*\;}
confirming that ultrametric diffusion with the particular barrier landscape \logheights\ defines a {\it fast scrambler}. 

\subsec{Instability Of The Infinite Entropy Limit}

\noindent

A distribution of the form \logheights\ is well-known to have a a special significance in the context of general ultrametric diffusion laws \refs\virasoro. In standard applications
in the theory of disordered media one usually considers linearly growing landscapes, with a law of the form $I_n \propto n + {\rm constant}$. These systems exhibit quite slow diffusion, interpolating between localization-type behavior and standard gaussian random walks (cf. \refs\virasoro). On the other hand, logarithmic landscapes with a barrier growth of the form
\eqn\genkla{
I_n = \alpha\,\log (n) + {\rm constant}
}
lead to the so-called Kohlrausch-law behavior, with a diffusion law that can be tuned to be {\it faster} than a gaussian random walk. An interesting quantity which is sensitive to the dynamical coefficient $\alpha$ in \genkla\ is the return walk  probability. Starting with an initial distribution completely localized at a single origin point $i=o$, i.e. $\rho_i (0) = \delta_{io}$, we consider the return probability after time $t$, which can be written quite explicitly,
\eqn\retp{
\rho_o (t) = {1\over S_*} + \sum_{r=1}^{n_*} {d \over (d+1)^{n_*-r+1}} \,e^{-\Gamma_r \,t}\;,}
for the present case of a regular tree with branching $b=d+1$.
For a landscape of the form \genkla\ the spectrum of quasi-normal modes reads  
\eqn\qnml{
\Gamma_r  \propto  {1\over (n_*-r+1)^\alpha} + d\sum_{k=n_* -r+1}^{n_*} {1\over k^\alpha}\;,}
up to an overall  $\CO(1)$ constant $c$. Inserting this expression back into \retp, after relabeling the index sum we obtain
\eqn\rtp{
\rho_o (t) = {1\over S_*} + \sum_{l=1}^{n_*} {d \over (d+1)^l} \exp\left(-c\,t\left({1\over l^\alpha} + d\sum_{k=l}^{n_*} {1\over k^\alpha}\right)\right)\;.
} 

The argument of the exponential in \rtp\ scales as $1/l^{\alpha-1}$ in the   $n_* \rightarrow \infty$ limit, provided $\alpha>1$. Otherwise, the random walk is not stable in the thermodynamical limit. 
Assuming then $\alpha >1$ we may estimate  the $n_* \rightarrow \infty$ limit  by saddle point approximation after converting the $l$ sum into an integral, obtaining 
\eqn\kl{
\rho_o (t)_{S_*\rightarrow \infty} \sim \exp\left(-c'\,t^{1/\alpha}\right)\;,}
the so-called stretched exponential or Kohlrausch law. 

Hence, we find a very satisfying result. Our {\it ansatz} \logheights, which was motivated by causality constraints in the near-horizon geometry, 
is exactly the critical ultrametric landscape in which the thermodynamic limit becomes impossible. Indeed, for $\alpha=1$ the exponents
in \rtp\ all diverge logarithmically as $S_*\rightarrow \infty$  and the random walk is completely spread up to infinity at any time $t>0$. 

The behavior of the autocorrelation function $\rho_o (t)$ is usually characterized in terms of the  {\it spectral dimension}, ${\tilde d}$, defined by  the long-time behavior as 
\eqn\spdim{
\rho_o (t) \sim t^{-{\tilde d} /2}\;.
}
With this definition, the Kohlrausch law \kl\ has an effectively infinite spectral dimension at late times, a behavior that is characteristic of fast scramblers and diffusion in expander graphs in general. 

Incidentally, the ballistic properties  can be codified in terms of \kl\ as well, since the number of points covered by the distribution in time $t$ is proportional to $1/\rho_o (t)$. The ultrametric radius of this `ball' is thus 
\eqn\bal{
r_{\rm eff} (t) \sim \log \left(1/\rho_o (t)\right) \sim t^{1/\alpha}\;.}
Therefore, we recover the expected ballistic behavior $r_{\rm eff} (t) \sim t$ precisely in the limit $\alpha \rightarrow 1^+$. 

We can view  the  $\alpha \rightarrow 1^+$ limit of \genkla\ as an analytic regularization of \logheights\  with the property of stabilizing the $S_* \rightarrow \infty$ limit and making contact with existing analysis in the theory of ultrametric diffusion. It should be stressed, however, that the present application to black-hole dynamics does not rely on the existence of a smooth infinite-entropy limit. 

\subsec{The Importance Of Being Decohered}

\noindent

Our entire discussion is based on a classical diffusion model,  defined on a peculiar substrate. It is interesting to ask for possible quantum generalizations of this set up. The direct reinterpretation of $W$ as a quantum Hamiltonian propagating an amplitude, rather than
a probability density, leads to the Schr\"odinger equation 
\eqn\sch{
i\pt_t \,\psi_i (t) = \sum_{j} W_{ij} \,\psi_j (t)\;.
}
Solutions with a real initial condition, $\psi_i (0) \in {\bf R}$,  can be expressed  in the form $\psi (t) = \rho(-it)$, where $\rho(t)$ solves the diffusion equation \diffe. A probability density would be obtained form $\psi(t)$ in the usual fashion, as $P_i (t) = |\psi_i (t)|^2$. The physical interpretation is that of a quantum probe hopping on an ultrametric lattice. It is easy to see that such system exhibits localization properties, corresponding to the fact that eigenvectors of $W_{ij}$ are indeed localized in `position space'.  Given a branch point $B$ at level $n$, the vector
$$
v_i (B) = {1\over (d+1)^n} \chi_i (B) - {1\over (d+1)^{n+1}} \chi_i (F_B) 
$$
is an eigenvector of $W_{ij}$, where $\chi_i (B)$ is the characteristic function of all B-descendants (equal to $1$ if $i$ descends from $B$ and equal to $0$ if $i$ does not descend from $B$). The point $F_B$ is the `father' or first ancestor of $B$. Hence, we see that $v_i (B)$ has a
degree of localization given by the characteristic function of the immediate ancestor (cf. \refs\ultradiff\ for detailed explanations). Initial components of the wave function with power on localized eigenvectors will remain so under time evolution. To be more explicit, consider the value of the amplitude at the origin, for a delta-function initial state $\psi_i (0) = \delta_{io}$, following from \rtp\ by simple analytic continuation:
$$
\psi_o (t) = {1\over S_*} + \sum_{r=1}^{n_*} {d \over (d+1)^{n_*-r+1}} \,e^{i\Gamma_r \,t}\;.
$$
The time-averaged quantum probability at the origin, 
$$
{\bar P}_o (t) = {1\over t} \int_0^t dt' \,|\psi_o (t')|^2
$$
approaches 
$$
{\bar P}_o (\infty) = {1\over S_*^2} + \sum_{r=1}^{n_*} {d^2 \over (d+1)^{2 (n_* -r+1)}} = {1\over S_*^2} + {d \over d+2} \left(1-{1\over S_*^2}\right)
$$
in the infinite-time limit. Hence, we see that the probability of remaining at the origin does not vanish in the limit $S_* \rightarrow \infty$, but rather approaches the constant $d/(d+2)$. At large but finite $S_*$ it remains well above the uniform probability, of order $1/S_*$, which would correspond to a delocalized state. Hence, we see that the ultrametric structure can only describe scrambling of probes after a suitable coarse-graining which must include decoherence effects. It is tempting to speculate that perhaps the interactions of the SH with the thermal atmosphere are enough to effect this required decoherence. 

These considerations bring into focus a very interesting toy model for the study of black hole scrambling (cf. Figure 2).  We can  consider a local  Q-bit model on a  cutoff expander graph such as the Cayley tree, as a minimal representation of the thermal atmosphere, in interaction with a non-local Q-bit model defined on the boundary of the tree,  mimicking the SH. The reduced density matrix of the SH is defined as the standard reduced state after tracing out the atmospheric degrees of freedom in the tree:
$$
\rho_{\rm SH} = {\rm Tr}_{\rm tree} |\Psi \rangle \langle \Psi |\;.
$$
 The degree of non-locality in the underlying boundary Q-bit model should be just right in order to recover the present ultra metric diffusion model after suitable decoherence reduces the full density matrix $\rho_{\rm SH}$ to the diagonal probability density $\rho_i(t) $ considered in the analysis of this paper. 
It is a very interesting open problem to determine if such a model can be constructed explicitly, and the interaction between boundary and bulk degrees of freedom be adjusted so that the required decoherence takes place. 

\bigskip
\centerline{\epsfxsize=0.6\hsize\epsfbox{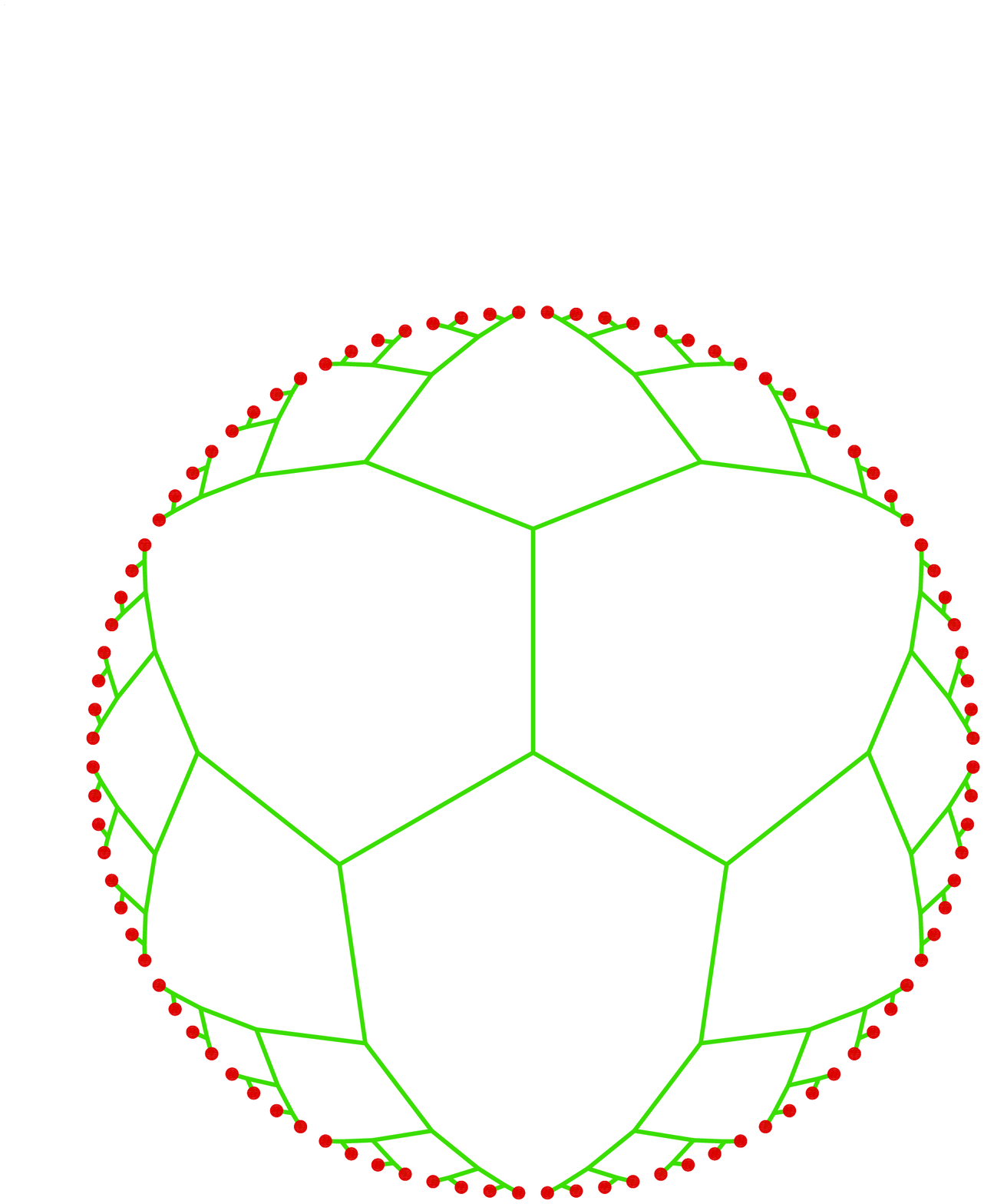}}
\noindent{\ninepoint\sl \baselineskip=2pt {\bf Figure 2:} {\ninerm
A discrete model encapsulating the scrambling properties of horizons consists of a local Q-bit model defined on the vertices of a cutoff tree, with non-local dynamics on the boundary sites (red dots), in such a way that the boundary state undergoes ultrametric diffusion after decoherence. 
}}
\bigskip

\newsec{Discussion}

\noindent

We have shown that non-local diffusion models on the stretched horizon, which inherit the causal-saturation properties of the near-horizon region are naturally fast scramblers with the characteristic $\log S_*$ time scale. We found this property to be equivalent to a particular type of ultrametric structure, lying at the critical line separating stable and unstable behavior in the large $S_*$ limit. To be more precise, the model is defined by approximating the transition rates between points on the SH by ultrametric distances in a discrete approximation of the near-horizon optical metric. This simple set up yields a fast scrambler with no strict stability in the thermodynamic limit $S_* 
\rightarrow \infty$. From the physical point of view, it is unclear if a smooth thermodynamic limit is a tight requirement. It is however interesting to note that a stable diffusion can be formally obtained by an analytic regularisation of the landscape structure, given by  \genkla,  in the limit $\alpha \rightarrow 1^+$ which, while not commuting with  $S_* \rightarrow \infty$, defines a stable model with ballistic scrambling if the thermodynamic limit is taken first in the regularised model. It would be interesting to see if more complicated diffusion laws, perhaps incorporating a degree of non-locality in the time variable, make the model less critical regarding the thermodynamic limit. 

We have also shown that the ultrametric structure, if relevant to the problem of black hole scrambling, must emerge as an effective construction after decoherence, since the hopping of quantum particles on an ultrametric structure leads to localization rather than scrambling. Such matters can be studied in an explicit toy model where the SH is modelled as a non-local Q-bit model in interaction with a `black hole atmosphere' furnished by a local Q-bit model defined on a cutoff tree.

Further work is needed to determine whether {\it ab initio} motivated models for stretched horizons (based for example on matrix models)
can be shown to exhibit the ultrametric structures uncovered in this work. It is plausible that these ideas could be of relevance for attempts at a direct proof along the lines of \refs\lashkari.

\bigskip{\bf Acknowledgements:} 
We whish to thank Auditya Sharma, Douglas Stanford and Lenny Susskind for discussions. 
This work  was partially supported by MEC and FEDER under a grants FPA2009-07908 and FPA2012-32828, the Spanish
Consolider-Ingenio 2010 Programme CPAN (CSD2007-00042), Comunidad Aut\'onoma de Madrid under grant HEPHACOS S2009/ESP-1473 and the 
spanish MINECO {\it Centro de excelencia Severo Ochoa Program} under grant SEV-2012-0249. The work of J.M.M was partially supported by  the Brazilian CNPq institution.

{\ninerm{
\listrefs
}}

\bye